\documentclass[twocolumn,prb,showpacs]{revtex4}

\usepackage{amssymb}
\usepackage{graphicx}
\usepackage{epsfig}
\usepackage{dcolumn}
\usepackage{bm}
\usepackage{amsmath}
\begin{document}
\title{Memristive model of amoeba's learning}
\author{Yuriy V. Pershin$^{1,2}$}\email{pershin@physics.sc.edu}
\author{Steven La Fontaine$^1$}
\author{Massimiliano Di Ventra$^{1,}$} \email{diventra@physics.ucsd.edu}
 \affiliation{$^1$Department of
Physics, University of California, San Diego, La Jolla, California
92093-0319 \\ $^2$Department of Physics and Astronomy and USC
Nanocenter, University of South Carolina, Columbia, SC, 29208}

\begin{abstract}
Recently, it was shown that the amoeba-like cell {\it Physarum
polycephalum} when exposed to a pattern of periodic environmental
changes learns and adapts its behavior in anticipation of the next
stimulus to come. Here we show that such behavior can be mapped into
the response of a simple electronic circuit consisting of an $LC$
contour and a memory-resistor (a memristor) to a train of voltage
pulses that mimic environment changes. We also identify a possible
biological origin of the memristive behavior in the cell. These
biological memory features are likely to occur in other unicellular
as well as multicellular organisms, albeit in different forms.
Therefore, the above memristive circuit model, which has learning
properties, is useful to better understand the origins of primitive
intelligence.
\end{abstract}


\maketitle

\section{Introduction}

Although it is a unicellular organism, {\it Physarum polycephalum}
displays remarkably intelligent abilities: it is able to solve mazes
\cite{litr1} and geometrical puzzles\cite{litr2}, control
robots\cite{litr3}, and may even be able to learn and recall past
events\cite{litr4}. According to Ref. \onlinecite{litr4}, when
exposed to three spikes of cold temperature and low humidity, set at
specific lengths of time and given at regular intervals, the {\it
Physarum} decreased its movement speed at the same time as the
shocks. However, after the spikes had stopped, the {\it Physarum}
still decreased its speed at the times when the spikes would have
occurred, effectively predicting the time of the spikes from the
pattern it had been given.  These "memory patterns" dissipate over
time, but a single spike in temperature can trigger the oscillations
again, provided it is within a certain time frame. This shows the
amoeba ``learned'' that a single temperature spike can be followed
by others at the given period.

At this point, the question may arise as to whether it is at all
appropriate to use the term ``learning'' to describe the amoeba's
response. Usually, when we discuss about learning in the animal
world, that term refers to a more complex behavior (for example,
classical conditioning and associative
memory~\cite{litr44,associatemem}) which is observed in higher
developed species such as mammals. In amoebas, this type of
classical conditioning has not been observed yet, and their
``learning'' is rather primitive. Irrespective, the adaptive
behavior of amoebas is astonishing, with direct experimental
evidence that amoebas can memorize sequences of periodic
environmental changes and recall past events. Therefore, in this
paper we will use the term ``learning'' only in the context of
recent experiments on this system~\cite{litr4}.

\begin{figure}[b]
 \begin{center}
\includegraphics[angle=0,width=5.0cm]{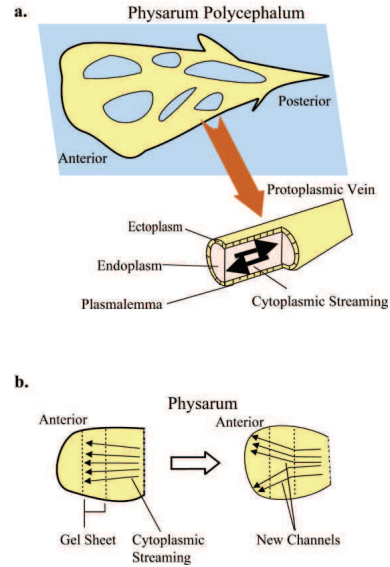}
\caption{\label{fig1}(Color online) View of {\it Physarum} during
movement.  (a) {\it Physarum} moves via shuttle streaming, a process
of periodic flow of the ectoplasm and endoplasm, with a greater net
flow towards the direction of movement, or the anterior end. The
ectoplasm contains radial and longitudinal actin-myosin fibers whose
oscillating contractions help produce the pressure gradient which
drives shuttle streaming\cite{litr5}. (b) During movement, the
anterior of {\it Physarum} may develop a sheet of gel that inhibits
streaming. However, when the pressure gradient builds to a certain
threshold, the gel can break down allowing for the formation of new
channels of flow. These channels may become new permanent veins for
streaming.}
 \end{center}
\end{figure}

A partial model to describe this behavior has been advanced in Ref.
\onlinecite{litr4} in terms of the ubiquitous biological
oscillators. According to this model, these internal oscillators
have a natural frequency, and may possibly deviate from that
frequency so that multiple oscillators can respond to the same
frequency.  The frequency of natural shocks excites one or more of
these oscillators, which could be the source of {\it Physarum}'s
ability to recognize patterns and predict events.  The model,
however, does not fully explain the memory response of the amoeba
and does not take into account the fact that, at a microscopic
level, other changes in the physiology of the organism may occur in
addition to the biological oscillators. These changes also occur
over a finite period of time and must be dependent on the state of
the system at previous times.

This last point is particularly important: it is in fact this
state-dependent feature which is likely to produce memory effects
rather than the excitation of biological oscillators. Instead, the
latter ones seem to control the rhythmical flow of the endoplasm
through protoplasmic veins\cite{litr5} shown schematically in Fig.
\ref{fig1}(a). As of now, it is not yet known how the memory in
amoebas is actually realized. However, we identify a possible
mechanism of amoeba's memory as described below. Irrespective of
whether this is the only mechanism leading to the observed response,
our model is in excellent agreement with the recent experimental
observations \cite{litr4} and can be used, in the form of
differential equations, to describe the amoeba's response to a
changing environment.

\begin{center}
\begin{table}
\begin{tabular}{| c | c | c | }
  \hline
  Component & Amoeba & Electronic circuit \\ \hline
  Control parameter & Temperature and  & Applied voltage \\
&humidity& \\ \hline
  Output signal & Velocity & Voltage \\  \hline
  Oscillator & Biochemical oscillators & $LC$ contour \\  \hline
   Memory element & Gel/sol interactions & Memristor \\  \hline
    Parameter con- & Veins and low- & Resistance of \\
taining information &viscosity channels & memristor \\
     \hline
\end{tabular}
\caption{Possible correspondence between learning of an amoeba and
an electronic circuit.} \label{tabl1}
\end{table}
\end{center}

Let us then consider mechanisms existing in amoebas which depend on
the state of the system and on its dynamical history thus
potentially giving rise to the observed memory response. These
mechanisms are as follows. The {\it Physarum} contains in its
interior a gel-sol solution.  The gel, present in the ectoplasm, is
more gelatinous than the less viscous sol, present in the endoplasm,
and the sol flows through the gel almost in the same way as water
through a sponge.  Now, the gel-sol solution is thixotropic, meaning
that the viscosity can change as a function of pressure. When the
amoeba is moving, the actin-myosin fibers present in the ectoplasm
contract radially and longitudinally, creating a pressure gradient
pushing the endoplasm in the direction of motion.  This gradient can
increase to the point in which it causes the gel to break down into
sol so that new low-viscosity channels form, which may even become
permanent pathways (see Fig. \ref{fig1}(b))\cite{litr6,litr7}.
Therefore, if the external temperature and humidity of the
environment are changed, the sol flow changes in a non-linear way. A
restoration of initial conditions upon change of the environment
thus requires time, and depends on the number and shape of the
formed low-viscosity channels. This mechanism is similar to the one
underlying the memory-resistance (memristor)
behaviour\cite{litr8,litr9,litr10,litr11,litr12,litr13,ournewarxivpaper}
of certain electronic devices\cite{litr10,litr11}. In these, the
variation of an external parameter (e.g., the voltage) creates new,
or modifies existing conducting channels thus altering the
resistance in a non-linear way. Therefore, it is natural to argue
that in the very same way that a memristor has its inherent memory,
the {\it Physarum} acquires a memory through the interactions of the
gel-sol solution, specifically through the formation of new,
low-viscosity channels, and perhaps by other related complex
interactions.

We now show this possible connection more clearly by presenting a
juxtaposition of the mechanisms behind the movement process of
{\it Physarum} with the operation of an equivalent electrical
circuit made of just the four basic passive electrical elements:
the resistor ($R$), capacitor ($C$), inductor ($L$), and
memristor\cite{litr8,litr14} ($M$). Indeed, we show that, like the
amoeba, this circuit can learn and predict subsequent signals. Due
to the complexity of biological systems, the present analogy will
be a simplification of what specifically happens during the
learning process. Nevertheless, it may be very useful in
understanding the origin of primitive intelligence in other
organisms as well.

\begin{figure}
 \begin{center}
\includegraphics[angle=0,width=5.5cm]{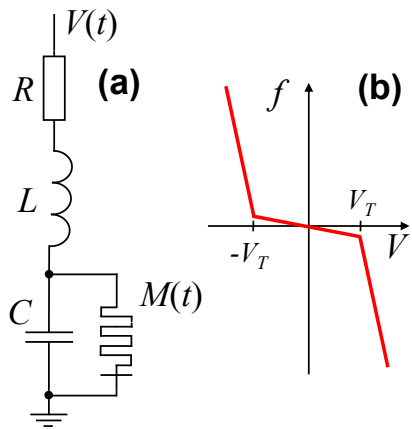}
\caption{\label{fig2}(Color online) Electronic circuit that models
amoeba's learning. (a) Schematic representation of the learning
circuit made of four fundamental two-terminal circuit elements:
resistor $R$, inductor $L$, capacitor $C$ and memristor $M$. (b)
Sketch of the selected memristor function  $f$ which depends on
the voltage applied to the memristor (more details are given in
the text). It is defined as $f(V)=-\beta V+0.5 (\beta-\alpha
)\left( |V+V_T|-|V-V_T| \right)$, where $\alpha$ and $\beta$ are
positive constants and $V_T$ is a threshold voltage.}
 \end{center}
\end{figure}

\section{Memristive circuit}

Let us then start by identifying the relation between the basic
circuit elements and the biological processes in the cell (see
Table \ref{tabl1}). In this respect the biological oscillators can
be simulated with the oscillations of an $LC$ contour. The
resistance $R$ describes the fact that there must be some signal
impedance and dissipation inside the amoeba, or else signals would
travel instantaneously and indefinitely. Finally the memristor $M$
summarizes the memory mechanisms we have described above, or any
other possible mechanism for memory. The temperature and humidity
that control the motion of the amoeba correspond to the external
voltage that controls the circuit. The response, namely the
amoeba's velocity, is nothing other than the voltage at the
memristor.

The electronic scheme that accomplishes the learning process in
response to a train of voltage pulses is shown in Fig.
\ref{fig2}(a). The capacitor and memristor are connected in
parallel. The main idea behind functioning of this scheme is to
use the internal state of memristor in order to store information
about the past and control oscillations in the $LC$ contour. In
particular, we use the model of a voltage-controlled memristor,
inspired by recent experiments\cite{litr11}, in which the
resistance of memristor $M$ can be changed between two limiting
values $M_1$ and $M_2$, $M_1<M_2$. In our scheme, $M$ increases as
a result of increased number of periodic stimuli and thus as a
result of learning.

The change of $M$ is described by the following equation

\begin{equation}
\frac{\textnormal{d}M}{\textnormal{d}t}=f\left( V_C\right) \left[
\theta\left( V_C\right)\theta\left( M-M_1\right) + \theta\left(
-V_C\right)\theta\left( M_2-M\right)\right], \label{eq1}
\end{equation}
where $f\left( V_C\right)$ is a function describing the change of
the memristor state, $V_C$ is the voltage applied to the memristor
(equal to the voltage drop on the capacitor) and $\theta\left(
...\right)$ is a step function. The expression in square brackets
guarantees that $M$ changes between $M_1$ and $M_2$. Here, we
assume that $f\left( V_C\right)$ consists of several linear
segments as shown in Fig. \ref{fig2}(b). This is the simplest
memristor model which takes into account the activation change of
the memristor state\cite{litr15}. In other words, the memristor
learns faster when $\left| V_C \right|>V_T$ and slower when
$\left| V_C \right|<V_T$, where $V_T$ is a threshold voltage.

The response of the circuit shown in Fig. \ref{fig2}(a) is
described by the following equations:

\begin{eqnarray}
V_C+L\dot I+IR=V(t), \label{eq2} \\ C\dot V_C+\frac{V_C}{M}=I,
\label{eq3}
\end{eqnarray}
where $V_C$ is the voltage on the capacitor, $I$ is the total
current and $V(t)$ is the applied voltage. Eq. (\ref{eq2}) simply
states that the applied voltage is equal to the sum of voltage
drops on each element of the circuit, and Eq. (\ref{eq3}) is the
Kirchhoff's current law at the point of connection of capacitor,
inductor and memristor. We solve Eqs. (\ref{eq2}) and (\ref{eq3})
together with Eq. (\ref{eq1}) numerically using initial conditions
close to a steady state and different $V(t)$.

\begin{widetext}

\begin{figure}[tb]
 \begin{center}
\includegraphics[angle=0,width=15cm]{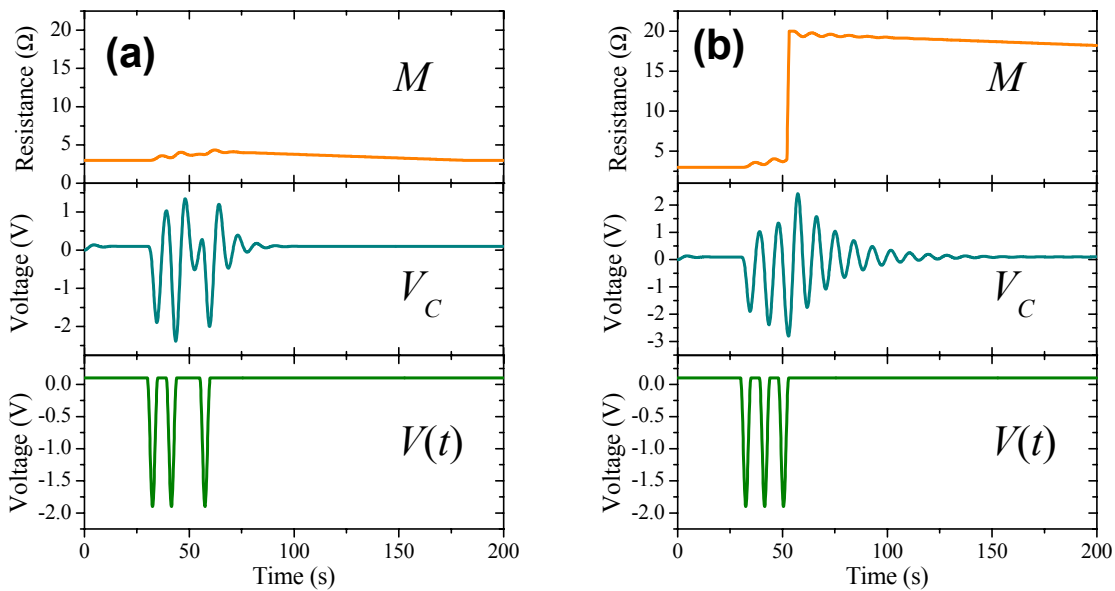}
\caption{\label{fig3}(Color online) Simulations of the circuit
response to applied pulse sequences. (a) Arbitrary pulse sequence.
(b) Learning pulse sequence. The calculations were made using the
following system parameters:  $R=1\Omega$, $L=2$H, $C=1$F,
$M_1=3\Omega$, $M_2=20\Omega$, $\alpha=0.1\Omega$/(Vs),
$\beta=100\Omega$/(Vs), and $V_T=2.5$V. The applied pulse sequence
was selected in the form $V(t)=V_F-V_p\sum\limits_i\left( \cos
\left(2\pi \left( t-t_i\right)/W_p \right)-1
\right)\theta\left(t-t_i \right)\theta\left(t_i+W_p-t \right)/2$,
where $V_F$ is the voltage corresponding to the standard
(favourable) conditions, $V_p$ is the pulse amplitude, $t_i$ is
the time of start and $W_p$ is the pulse width. In the
calculations, we used $V_F=0.1$V, $V_p=2$V, and $W_p=5$s. The
$t_i$'s can be identified from the figure.}
 \end{center}
\end{figure}

\end{widetext}

\section{Results and discussion}
In our circuit scheme, a
favorable (standard) environmental condition  corresponds to a
positive applied voltage and unfavorable condition to a negative
applied voltage. If the favorable condition is applied for a long
period of time, then, during this period of time, a positive
voltage is applied to the memristor. According to Eq. (\ref{eq1})
and $f\left( V_C\right)$ as in Fig. \ref{fig2}(b), the latter
switches into the low resistance state $M_1$. In this case, the
$LC$ contour is damped and excited oscillations decay fast. Fig.
\ref{fig3}(a) demonstrates that when a non-periodic sequence of
pulses is applied to the scheme, the circuit learning ability (or
change of $M$) is small and oscillations in the contour are
strongly damped. In the opposite case, when we apply periodic
pulses with a frequency close to the $LC$ contour's resonant
frequency, the change of $M$ is much more pronounced, the
memristor switches into its higher resistance state and, since the
$LC$ contour becomes less damped, oscillations in the contour
survive for a longer time. Such behavior is related to the fact
that during the application of resonant pulses the amplitude of
voltage oscillations on the capacitor increases with each pulse
and at some point exceeds $V_T$ in amplitude. As a consequence,
the memristor learns fast and its state changes significantly.

\begin{figure}[tb]
 \begin{center}
\includegraphics[angle=0,width=8.0cm]{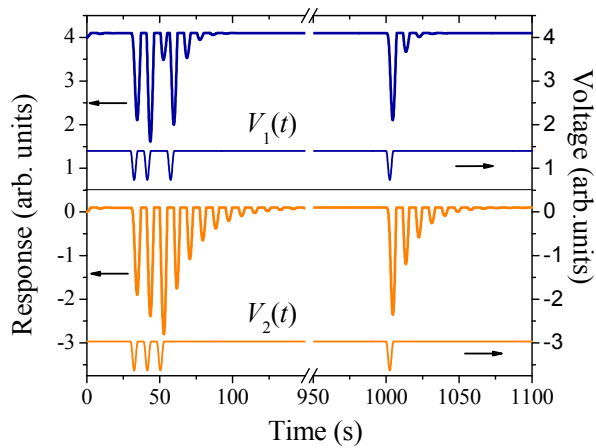}
\caption{\label{fig4}(Color online) Modeling of the spontaneous
in-phase slowdown responses. This plot demonstrates that stronger
and longer lasting responses for both spontaneous in-phase
slowdown and spontaneous in-phase slowdown after one disappearance
of the stimulus are observed only when the circuit was previously
``trained'' by a periodic sequence of three equally spaced pulses
as present in $V_2(t)$. The applied voltage $V_1(t)$ is irregular
and thus three first pulses do not ``train'' the circuit to learn.
The simulation parameters are as in Fig. \ref{fig3}. The lines
were displaced vertically for clarity. The arrows are used to
define the correct vertical axis for each line.}
 \end{center}
\end{figure}

Fig. \ref{fig4} shows simulations of spontaneous in-phase slowdown
(SPS) and SPS after one disappearance (SPSD) events as in the
experiments with the amoeba\cite{litr4} (a description of these
experiments is given below). In these simulations we use the
scheme described above with the only restriction that the response
signal can not exceed a certain value, which in our particular
calculations is selected to be equal to the voltage corresponding
to standard (favorable) conditions $V_F$. This signal is selected
to be equal to $V_C$ if $V_C<V_F$ and to $V_F$ if
$V_C>V_F$.\cite{prec}

In SPS experiments, the amoeba was exposed to three intervals of
unfavorable conditions (namely, low temperature and humidity). Each
time, the locomotion speed decreased. After that, standard
(favorable) conditions were applied. However, the movement of the
amoeba has been found to slow spontaneously when the next intervals
of unfavorable conditions would have occurred. Exactly the same
behavior is found in our circuit model. This is shown in Fig.
\ref{fig4}, where a regular pulse sequence $V_2(t)$ is applied. It
is clearly seen that in response to the application of three regular
pulses, the response signal decreases each time these pulses are
applied as well as at subsequent times when following pulses would
have occured. The opposite behavior is demonstrated when the three
training pulses are not periodic as shown in Fig. \ref{fig4} with an
irregular pulse sequence $V_1(t)$. In this case, the anticipated
response is significantly smaller.

Application of the fourth pulse in Fig. \ref{fig4} corresponds to
conditions of SPSD experiments\cite{litr4} in which the anticipated
slowdown after a single unfavorable condition was observed only
among previously trained organisms. Again, we can see a striking
similarity with the experimental results on amoeba's learning: in
the case of a pulse sequence with the first three non-periodic
pulses ($V_1(t)$), the subsequent application of a single pulse does
not result in a significant anticipated slow-down after the pulse.
On the other hand, a ``trained'' circuit (by applying the $V_2(t)$
pulse sequence) manifests several well-defined slowdown events after
the fourth pulse.

\begin{figure}[tb]
 \begin{center}
\includegraphics[angle=0,width=7.0cm]{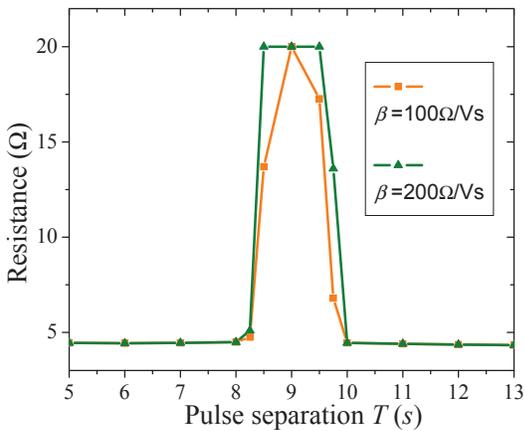}
\caption{\label{fig5}(Color online) Resistance of memristor
calculated at two different values of the parameter $\beta$ in the function $f(V_C)$ as a
function of time interval $\tau$ between the pulses. Here, we plot
the value of the memristance $M$ right after the last (third)
pulse in the learning sequence. The memristor state significantly
changes when pulse period is close to the $LC$ contour frequency.
The simulation parameters are as in Fig. \ref{fig3}.}
 \end{center}
\end{figure}

Finally, we plot in Fig. \ref{fig5} the final state of the
memristor just after the application of three learning pulses as a
function of time interval between the pulses. It can be noticed
that learning occurs in the interval 8s$<\tau <$10s, which is
close to the $LC$ contour time sequence. Also, the larger the rate
at which the memristor learns (as exemplified by the parameter
$\beta$ in the function $f(V_C)$) the more defined is the time
interval for learning. To some extent, this plot resembles Fig. 3 in
Ref. [\onlinecite{litr4}] for amoeba's learning. A qualitative
description of the dependence shown in Fig. \ref{fig5} is the
following. It is well known that the amplitude of oscillations
excited in a $LC$ contour decreases when the applied voltage
frequency moves away from the resonant frequency. In particular,
if we consider just a simple $LC$ contour (as in Fig.
\ref{fig2}(a) but without $R$ and $M$) driven by $V(t)=V_0
\cos(\omega t)$, then the amplitude of steady-state voltage
oscillations on the capacitor is
\begin{equation}
V_C^0=\left| \frac{V_0}{1-LC\omega^2} \right|. \label{simpleLC}
\end{equation}
If we now assume a memristor connected in parallel with the
capacitor in such a contour, to first approximation, $M$ can be
significantly changed only if $V_C$ exceeds the memristor threshold
$V_T$. This occurs when, according to Eq. (\ref{simpleLC}), the
frequency $\omega$ of the applied field is close to the resonance
frequency $1/\sqrt{LC}$. Fig. \ref{fig5} clearly demonstrates that
the range of periods of applied oscillations leading to significant
changes in $M$ is distributed around the period of resonant
oscillations in the contour, which in the present case is $T=2 \pi
\sqrt{LC}=8.9$s.

We have demonstrated that a simple electronic circuit can be used
to model the results found in Ref. \onlinecite{litr4}, which show
that an amoeba migrating across a narrow lane can learn the period
of temperature shocks, and predict when future shocks would occur.
As the amoeba migrates in normal conditions, new veins form in the
amoeba as a natural process of movement, and old ones decay into
the tail of the amoeba (see Fig \ref{fig1}). As a result, the
amoeba's internal structure dynamically changes with vein
formation and degradation. The vein formation is pressure
dependent, and the pressure is a result of the constriction of the
sol by the actin fibers in the gel, with both radial and
longitudinal contractions. This contraction is rhythmic and
periodic with ATP and Ca$^{2+}$ levels, and possibly with other
biological compounds as well. Thus, in amoeba's motility, it is
the internal oscillations of actin fiber contractions that create
the pressure gradients that control the formation of new veins and
the decay of old ones; movement entails change in the vein
structure, whether or not an outside periodic stimulus is present.
The circuit in Fig. \ref{fig2} can be seen as describing the
process of vein formation, whereby LC oscillations generate the
voltage threshold to change the state of the memristor. In the
amoeba, the formation of veins means that sol can flow more
easily; in our circuit, a high resistance state in the memristor
means that the input signal is more conserved, so that current can
flow more easily.

Ref. \onlinecite{litr4} explains that a periodic stimulus may in
fact link certain biological oscillators together, producing a
strong rhythmic response to that stimulus. Qualitatively speaking,
assuming these oscillators link together locally in the amoeba, a
strong local response would aid vein formation in that local area so
that the sol flow caused by those oscillators is less damped than
sol flow in veins that resulted from a normal or weak oscillatory
response. To reiterate, a periodic stimulus will produce a stronger
periodic response, assuming that biological oscillators link
together, resulting in stronger vein formation, as a consequence of
which the oscillatory response can be more conserved than usual.  We
argue that therein lies the memory of the amoeba: the passive
``decision'' to conserve the strongest output signal produced by a given
input signal. This is the effect illustrated by our circuit; a
strong LC response is conserved by a high memristor value.  Vein
structures are changed over time by the movement of the amoeba in
response to any given stimulus; however, it seems that a periodic
stimulus produces the strongest response that leads to memory.
Clearly, due to the complexity of the biological problem at hand, we
cannot exclude that other mechanisms are at play in producing the
observed amoeba's memory.

\section{Conclusions}

In conclusion, we have presented an electronic circuit model to
describe the amoeba's ability to recognize patterns and predict
events. This model contains a memristive element and simulates the
mechanisms of biological memory that possibly occur in the
protoplasm of the {\it Physarum}, which produce a sol flow that
depends on the history and state of the system. A collection of
circuits as those presented here (but with different resonant
frequencies), or, possibly, a single circuit with the replacement
of the capacitor and/or inductor with the newly introduced
memory-capacitor (memcapacitor) and memory-inductor (meminductor)
\cite{litr9a}, closely simulates the experimentally observed
learning ability of the amoeba \cite{litr4} including the memory
of period, and provides a dynamic picture of the memory mechanism
in this unicellular organism. It is worth noting that the proposed
electronic circuit is made only of passive elements and can be
realized in the laboratory. It may thus find applications in
electronic applications that require ``circuit learning'' or
pattern recognition. Finally, this model may be extended to
multiple learning elements and may thus find application in neural
networks and guide us in understanding the origins of primitive
intelligence and adaptive behavior.

\section*{Acknowledgements} Y.V.P and M.D. acknowledge partial
support from the NSF grant No. DMR-0802830. S.L. acknowledges
support from the University of California, San Diego, CAMP Science
Program for undergraduate students.


\begin{thebibliography}{}
\bibitem{litr14} In this paper we use the word memristor in a broad sense to
include both perfect memristors \cite{litr8} and memristive
systems\cite{litr9}.
\bibitem{litr15} This model is inspired by a recent experimental paper (Ref.~\onlinecite{litr10}), where
it is shown that the memristor state changes fast at high applied
voltages, and varies little at low voltages.
\bibitem{prec} Electronically, such a response signal can be
obtained using a resistor connected from one side to the junction
of the inductor, capacitor and memristor in Fig. \ref{fig2}(a) and
a diode connected between another side of the resistor and a power
source at $V_F$ voltage.
\end{thebibliography}
\end{document}